\documentclass[epj]{webofc}
\usepackage[utf8]{inputenc}
\usepackage[varg]{txfonts}   
\usepackage{booktabs}
\usepackage{xcolor}
\usepackage{tikz}
\usepackage{mwe}
\definecolor{darkred}{rgb}{0.4,0.0,0.0}
\definecolor{darkgreen}{rgb}{0.0,0.4,0.0}
\definecolor{darkblue}{rgb}{0.0,0.0,0.4}
\usepackage[bookmarks,linktocpage,colorlinks,
    linkcolor = darkred,
    urlcolor  = darkblue,
    citecolor = darkgreen]{hyperref}
\pdfpageattr{/Group << /S /Transparency /I true /CS /DeviceRGB>>}
\usepackage{subfigure}
\usepackage{graphicx,wrapfig}
\usepackage{macros}
\usepackage{multirow}
\usepackage{amssymb,amsmath}
\usepackage{booktabs,dcolumn}
\usepackage{grffile}         
\usepackage{lipsum}
\newcolumntype{C}{>{$}c<{$}} 
\wocname{EPJ Web of Conferences}
\woctitle{Lattice2017}
\begin{document}
%
\selectlanguage{english}
\hfill\parbox{14mm}{\texttt{
MS-TP-17-19
}}
\title{%
\vspace{-0.5cm}
Leptonic decay constants for D-mesons from 3-flavour CLS \\
ensembles\thanks{%
Talk at the 35th International Symposium on Lattice Field Theory
(LATTICE 2017), 18-24 June 2017, Granada, Spain
}\\\\
\includegraphics[width=2.125cm]{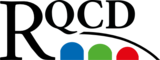}%
\hfill%
\includegraphics[width=2.5cm]{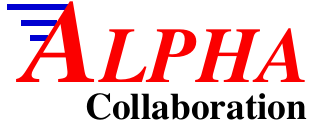}%
}
\author{%
\firstname{Sara} \lastname{Collins}\inst{1} \and
\firstname{Kevin} \lastname{Eckert}\inst{2}\fnsep\thanks{Speaker, \email{k.eckert@uni-muenster.de}} \and
\firstname{Jochen} \lastname{Heitger}\inst{2} \and
\firstname{Stefan} \lastname{Hofmann}\inst{1} \and
\firstname{Wolfgang}  \lastname{S\"oldner}\inst{1}
}
\institute{%
Universit\"at Regensburg, Institut f\"ur Theoretische Physik, 93040 Regensburg,
Germany
\and
Westf\"alische Wilhelms-Universit\"at M\"unster, Institut f\"ur Theoretische
Physik, Wilhelm-Klemm-Straße 9, 48149 M\"unster, Germany
}
\abstract{%
We report on the status of an ongoing effort by the RQCD and ALPHA Collaborations,
aimed at determining leptonic decay constants of charmed mesons. Our analysis is
based on large-volume ensembles generated within the CLS effort, employing
$N_{\rm f}=2+1$ non-perturbatively $\mathcal{O}(a)$ improved Wilson quarks, tree-level
Symanzik-improved  gauge action and open boundary conditions. The ensembles
cover lattice spacings from $a\approx0.09\,\fm$ to $a\approx0.05\,\fm$, with pion masses varied
from $420$ to $200\,\MeV$. To extrapolate to the physical masses, we follow both
the $(2m_{\rm l}+m_{\rm s})=const.$ and the $m_{\rm s}=const.$ lines in parameter space.
}
\maketitle
\section{Introduction and computational setup}\label{intro}
\noindent
The pseudoscalar decay constants $\fD$ and $\fDs$ encode the
QCD contributions in leptonic decays of $\mathrm{D}$- and $\mathrm{D}_{\mathrm{s}}$-mesons,
respectively. Theory input on the decay constants in
conjunction with experimental data allows the CKM matrix  elements $V_{\rm cd}$
and $V_{\rm cs}$ to be tightly constrained. Apart from being an important test
of the Standard Model, hints of new physics may be discovered in the
charm sector. For an overview of lattice QCD results, see Ref.~\cite{Aoki:2016frl}.
An update on last year's report~\cite{Collins:2016proc} and our ongoing effort on
the computation of $f_{\mathrm{D}_{(\rm s)}}$ is presented in the following.

We utilize ensembles generated within the Coordinated Lattice Simulations (CLS)
effort, with $N_{\rm f} = 2 + 1$ non-perturbatively $\mathcal{O}(a)$ improved
Wilson-Sheikholeslami-Wohlert (clover) fermions and tree-level improved
L\"uscher-Weisz gauge action, employing the \texttt{openQCD}~\cite{openQCD} open-source
software package. Open boundary conditions in temporal direction are used in
order to avoid topological freezing, making the use of very fine lattice
spacings of $a\approx0.0854 - 0.039\,\fm$ ($\beta=3.4 - 3.85$) feasible~\cite{Luscher:2011kk,Luscher:2012av}. We also
employ twisted-mass reweighting for light quarks~\cite{Luscher:2008tw}, to prevent instabilities
resulting from accidental near-zero eigenmodes of the Dirac operator, and
rational approximation for the strange quark with appropriate reweighting.
Expectation values of physical observables are therefore obtained by
\begin{equation}
 \langle O \rangle=
 \frac{\langle OW_0W_1\rangle_W}{\langle W_0W_1\rangle_W}.
\end{equation}
The twisted-mass~($W_0$) and rational approximation
reweighting~($W_1$) factors are defined in
Ref.~\cite{Bruno:2014jqa} (Eqs. (3.2) and (3.5), respectively).

Apart from additional ensembles for $\beta=3.4$ and $\beta=3.55$, our analysis
now also includes ensembles for the lattice spacings with $\beta=3.46$, $\beta=3.7$ and
$\beta=3.85$. We follow two lines in the light and
strange quark mass plane: (i) The average lattice quark
mass~($\overline{m}=\left(2\ml+\ms\right)/3$) is kept fixed such that the sum
of the renormalized quark masses is constant up to $\mathcal{O}(a)$ effects
(ensembles available for all $\beta$ values). (ii) The renormalized
strange quark mass is kept constant, again up to $\mathcal{O}(a)$ effects
(ensembles available for $\beta=3.4$ and $\beta=3.55$). For details on how an
almost constant renormalized strange quark mass was achieved, see Ref.~\cite{Bali:2016umi}.
We only mention that the vector Ward identity masses are defined as
\begin{equation}
m_{{\rm q}={\rm l(ight)},{\rm s}}=
(1/\kappa_{\rm q}-1/\kappa_{\rm crit})/(2a),
\end{equation}
where $\kappa_{\rm crit}$ is the critical hopping parameter value
at which the axial Ward identity (i.e., PCAC) quark mass in the symmetric limit, $\ml=\ms$,
vanishes.

\begin{figure}[h]
  \centering
  \includegraphics[width=0.8\textwidth]{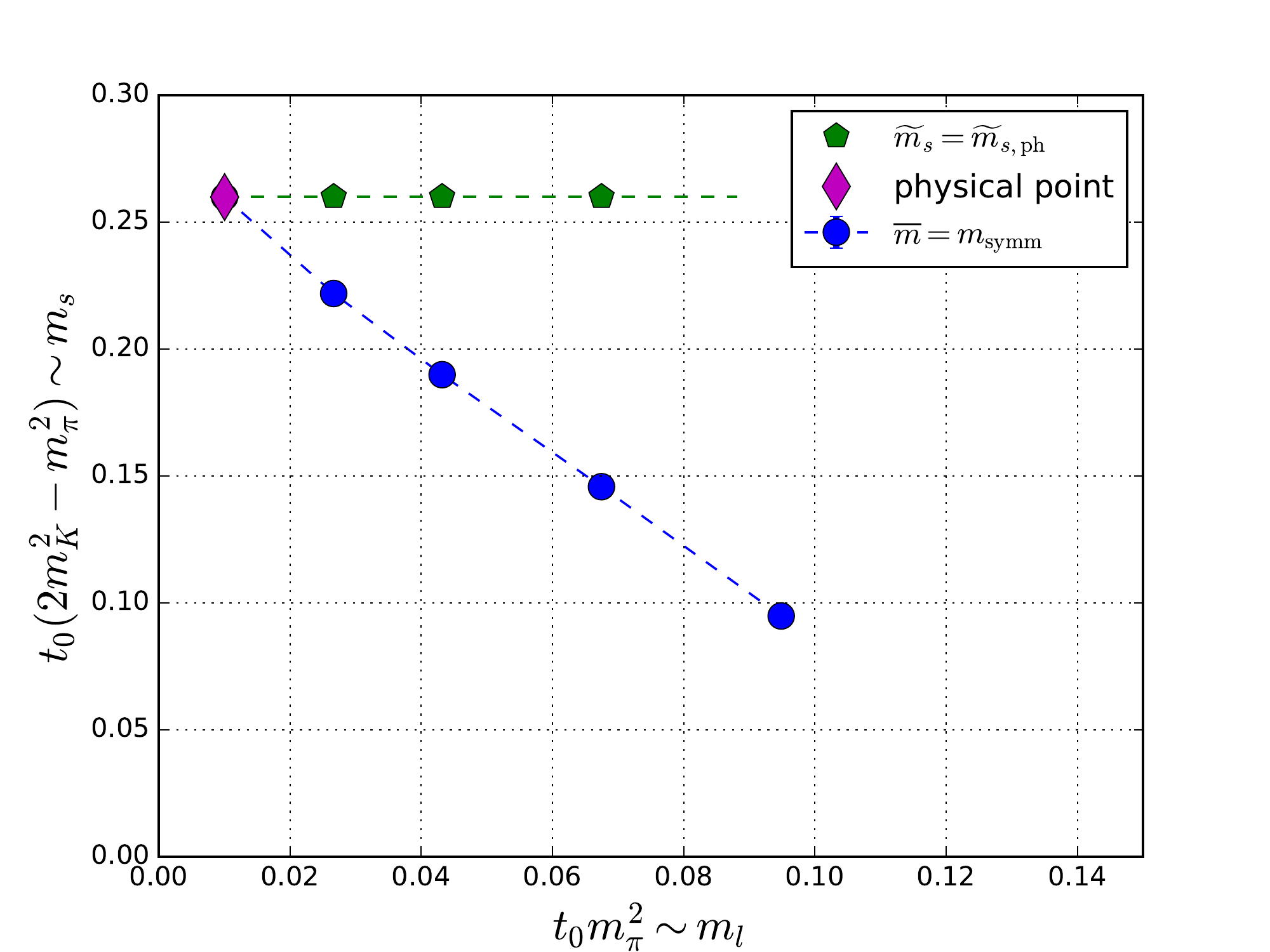}
  \caption{The light and strange quark masses realized for
    $\beta=3.4$ lattices as indicated by the square of the pion mass
    versus the kaon-pion mass difference,
    $2m_K^2-m_\pi^2$, in units of the gradient flow scale parameter $t_0$~\cite{Borsanyi:2012zs,Bruno:2014jqa,Bali:2016umi}. Pion masses here vary from $422$ to
    $223\,\MeV$.}
  \label{fig:QuarkMassPlane}
\end{figure}

The two lines are referred to as $\overline{m}=const.$ line (first proposed in~\cite{Bietenholz:2010jr})
and $\hat{m}_{\rm s}=const.$ line, respectively, and exemplarily visualized in
Fig.~\ref{fig:QuarkMassPlane} for the ensembles with $\beta=3.4$. Note that
the figure displays the PCAC strange quark mass $\tilde{m}_{\rm s}$,
for which the tuning to constant values was actually done but which corresponds to constant $\hat{m}_{\rm s}$ up to very small corrections.
Having two lines in the quark mass plane
available enables us to tightly constrain the chiral extrapolation, by
enforcing extrapolations along both lines to intersect at the physical point.
For further details concerning the computational setup see Refs.~\cite{Bruno:2014jqa,Collins:2016proc,Bali:2016umi,Bruno:2016plf}.

\begin{table}
{
\footnotesize
\begin{tabular}{ccllcccccc}
\toprule
trajectory&ensemble&\multicolumn{1}{c}{$\kapl$}&\multicolumn{1}{c}{$\kaps$}&$Lm_{\pi}$&$N_t\times N_s^3$&$\frac{m_{\pi}}{\textmd{MeV}}$&
$\frac{m_K}{\textmd{MeV}}$&$N_{\mathrm{MD}}$\\
\midrule
\multicolumn{9}{c}{$\beta=3.4$ [$a= 0.0854(15)\,\textmd{fm}$], \hspace{5pt} $\sqrt{8t_0}/a=4.852(7)$}\\
\midrule
\multirow{3}{*}{$\overline{m}=m_{\mathrm{sym}}$}
  &H101   &0.13675962 &0.13675962    & 5.8 & $96\times 32^3$  &
  422 & 422 &  8000\\
  &H102   &0.136865   &0.136549339   & 4.9 & $96\times 32^3$  &
  356 & 442 &  7988\\
  &H105   &0.13697    &0.13634079    & 3.9 & $96\times 32^3$  &
  282 & 467 &  11332\\
\hline
\multirow{2}{*}{$\hat{m}_{\rm s}=\hat{m}_{\rm s}^{\rm (phys)}$}
  &H107&0.136945665908&0.136203165143 & 5.1 & $96\times 32^3$&
  368&549&6256\\
  &H106&0.137015570024&0.136148704478 & 3.8 & $96\times 32^3$&
  272&519&6212\\
\midrule
\multicolumn{9}{c}{$\beta=3.46$ [$a \approx 0.077\,\textmd{fm}$], \hspace{5pt} $\sqrt{8t_0}/a \approx 5.43$}\\
\midrule
\multirow{2}{*}{$\overline{m}=m_{\mathrm{sym}}$}
  & S400         & 0.136984  & 0.136702387  & 4.3 & $128\times 32^3$  &
  347 & 437 &  6968\\
  & N401         & 0.1370616 & 0.1365480771 & 5.3 & $128\times 48^3$  &
  282 & 456 &  4400\\
\midrule
\multicolumn{9}{c}{$\beta=3.55$ [$a= 0.0644(11)\,\textmd{fm}$], \hspace{5pt} $\sqrt{8t_0}/a=6.433(6)$}\\
\midrule
\multirow{5}{*}{$\overline{m}=m_{\mathrm{sym}}$}
  & H200 & 0.137    & 0.137       & 4.4& $96\times 32^3$ &
  418 & 418 & 8000\\
  & N202 & 0.137    & 0.137       & 6.4& $128\times 48^3$ &
  410 & 410 & 3536\\
  & N203 & 0.13708  & 0.136840284 & 5.4& $128\times 48^3$ &
  345 & 441 & 6172\\
  & N200 & 0.13714  & 0.13672086  & 4.4& $128\times 48^3$ &
  283 & 461 & 6800\\
  & D200 & 0.1372   & 0.136601748 & 4.2& $128\times 64^3$&
  199 & 479 & 4000\\
\hline
\multirow{3}{*}{$\hat{m}_{\rm s}=\hat{m}_{\rm s}^{\rm (phys)}$}
  & N204 & 0.137112   & 0.136575049 & 5.6& $128\times 48^3$&
  351 & 544 & 2000\\
  & N201 & 0.13715968 & 0.136561319 & 4.5& $128\times 48^3$&
  284 & 522 & 6000\\
  & D201 & 0.137207   & 0.136546436 & 4.1& $128\times 64^3$&
  198 & 499 & 4312\\
\midrule
\multicolumn{9}{c}{$\beta=3.7$ [$a \approx 0.05\,\textmd{fm}$], \hspace{5pt} $\sqrt{8t_0}/a \approx 8.30$}\\
\midrule
\multirow{2}{*}{$\overline{m}=m_{\mathrm{sym}}$}
  & N300 & 0.137    & 0.137        & 5.1 & $128\times 48^3$ &
  418 & 418 & 8188\\
  & J303 & 0.137123 & 0.1367546608 & 4.2 & $192\times 64^3$ &
  257 & 473 & 2536\\
\midrule
\multicolumn{9}{c}{$\beta=3.85$ [$a \approx 0.0395\,\textmd{fm}$], \hspace{5pt} $\sqrt{8t_0}/a \approx 10.6$}\\
\midrule
\multirow{1}{*}{$\overline{m}=m_{\mathrm{sym}}$}
  & J500 & 0.136852    & 0.136852      & 5.2 & $192\times 64^3$ &
  404 & 404 & 3368\\
\bottomrule
\end{tabular} }
\caption{Details of the ensembles analyzed so far for the two
  trajectories to the physical point, keeping $\overline{m}$ fixed to
  the value at the symmetric point~($\overline{m}=m_{\rm sym}$) and
  keeping the renormalized strange quark mass~($\hat{m}_{\rm s}$) approximately equal to
  the physical value~($\hat{m}_{\rm s}^{\rm (phys)}$). The light and strange
  quark hopping parameters are denoted $\kapl$ and $\kaps$,
  respectively. The lattice volumes $N_t\times N_s^3$,
  the pion~($m_\pi$) and kaon~($m_K$) masses and the statistics given by
  the number of molecular dynamics units~($N_{\rm MD}$) are also indicated.}
\label{tab:CLSensembles}
\end{table}

\section{Observables}\label{sec-2}
\noindent
The pseudoscalar decay constants $\fD$ and $\fDs$ are the low-energy QCD
contributions to the leptonic decays of $\mathrm{D}$- and
$\mathrm{D}_{\mathrm{s}}$-mesons. They are
defined as the matrix elements
\begin{equation}
 \left\langle 0\left\vert A^{\rm qc}_\mu\right\vert\mathrm{D}_{\rm q}(p)\right\rangle=
 \mathrm{i} f_{\mathrm{D}_{\rm q}} p_{\mu},
 \label{eq:Fps}
\end{equation}
where $A^{\rm qc}_{\mu}=\overline{q}\gamma_\mu\gamma_5c$ is the axial vector
current for quark flavours ${\rm q}={\rm l(ight)},{\rm s}$ and  $\vert \mathrm{D}_{\rm q}(p)\rangle$
is a pseudoscalar meson state at momentum $p$ with quantum numbers corresponding to the
$\rm D$- (for ${\rm q}={\rm l}$) or $\rm D_{\rm s}$-meson (for ${\rm q}={\rm s}$).
We make use of the pseudoscalar operator
$P^{\rm qc}=\overline{q}\gamma_5c$ to remove $\mathcal{O}(a)$ discretization
effects from the axial operator, obtaining the improved axial current
\begin{equation}
 A^{\rm qc,I}_{\mu}=
 A^{\rm qc}_{\mu}
 +a\ca\frac{1}{2}\left(\partial_\mu+\partial^*_\mu\right)P^{\rm qc},
 \label{eq:CaImprove}
\end{equation}
with the standard notation for lattice forward and backward derivatives.
Renormalization then proceeds via
\begin{equation}
 \left(A^{\rm qc,I}_{\mu}\right)_{\rm R}=
 \za\left[1+a\left(\ba m_{\rm qc}+3\tilde{b}_{\rm A}\overline{m}\right)\right]
 A^{\rm qc,I}_{\mu} +{\cal O}(a^2).
 \label{eq:RenImp}
\end{equation}
Here, $m_{\rm qc}$ and $\overline{m}$ denote the bare vector Ward identity quark mass
combinations
\begin{equation}
 m_{\rm qc}=
 \frac{1}{2}\left(\mq+\mc\right), \qquad \overline{m}=\frac{1}{3} \left( \ms+2\ml \right).
 \label{vectorWardIdentity}
\end{equation}
The improvement coefficients $c_{\rm A}$ and $Z_{\rm A}$ have been calculated
non-perturbatively in Refs~\cite{Bulava:2015bxa,Bulava:2016ktf}.
$b_{\rm A}$ has been determined non-perturbatively in Ref~\cite{Korcyl:2016ugy}
and in a preliminary analysis the same authors find $\tilde{b}_{\rm A}$
to be consistent with zero. Since $\overline{m}$ includes only the light and
strange quark, the mass dependent corrections in Eq.~(\ref{eq:RenImp}) are dominated
by the charmed mass term $m_{\rm qc}$. Therefore we neglect the
term proportional to $\tilde{b}_{\rm A}\overline{m}$ in our analysis.

In order to extract the matrix elements of Eq.~(\ref{eq:Fps}), we evaluate
the two-point functions
\begin{eqnarray}
 C_{\mathrm{A}}(x_0, y_0)=
 -\frac{a^6}{L^3}\sum_{\vec{x},\vec{y}}
 \langle A_4^{\rm qc,I}(x)\left(P^{\rm qc}(y)\right)^\dagger\rangle, \qquad
 C_{\rm P}(x_0, y_0)=
 -\frac{a^6}{L^3}\sum_{\vec{x},\vec{y}}
 \langle P^{\rm qc}(x)\left(P^{\rm qc}(y)\right)^\dagger\rangle,
 \label{eq:corrs}
\end{eqnarray}
at zero momentum, with $y_0$ being the timeslice of the source insertion and $x_0$ that of the
sink. Starting from the spectral decomposition of the two-point functions
\begin{equation}
  C_{\rm X}(x_0,y_0) = \sum_{i=1}^{\infty} \; c_{\mathrm{X},i} \mathrm{e}^{-E_i (x_0-y_0)} \;\; \text{with} \;\; E_1=m_{\mathrm{D}_{\rm q}} \;,\; E_{i\geq2}:\text{excited states}\;,\;\mathrm{X} = \mathrm{P},\mathrm{A} \;,
  \label{spectral}
\end{equation}
it can be shown~\cite{Bruno:2016plf} that for large time separations $x_0-y_0$ the correlators behave as
\begin{align}
  \label{eq:corrs_exp}
  \nonumber C_{\mathrm{A}}(x_0,y_0)&\approx\frac{f^{\rm bare}_{\rm qc}}{2}\,A(y_0)\,\mathrm{e}^{-m_{\mathrm{D}_{\rm q}}(x_0-y_0)} \\
  \nonumber &\equiv c_{\mathrm{A},1}(y_0)\mathrm{e}^{-m_{\mathrm{D}_{\rm q}}(x_0-y_0)},
  \\
  \nonumber C_{\rm P}(x_0,y_0)&\approx\frac{\left|A(y_0)\right|^2}{2m_{\mathrm{D}_{\rm q}}}\,
  \mathrm{e}^{-m_{\mathrm{D}_{\rm q}}\left(x_0-y_0\right)} \\
  &\equiv c_{\mathrm{P},1}(y_0)\mathrm{e}^{-m_{\mathrm{D}_{\rm q}}(x_0-y_0)},
\end{align}
where the bare decay constant $f^{\rm bare}_{\rm qc}$ corresponds to $\langle 0 | A_4^{\rm qc,I} | \mathrm{D}_{\rm q} \rangle / m_{\mathrm{D}_{\rm q}}$,
while $A(y_0)$ encodes the matrix element $\langle 0\left|P^{\rm qc}\right|\mathrm{D}_{\rm q}\rangle$
plus possible boundary contaminations at the source position.

\section{Analysis details}\label{sec-3}
\noindent
Elaborating on our earlier status report in~\cite{Collins:2016proc}, we describe
our current analysis setup in this section. It features point-to-all propagators with
Wuppertal smearing~\cite{Gusken:1989ad,Gusken:1989qx} and APE-smoothed
links~\cite{Falcioni:1984ei} for the pseudoscalar source and sink operator $P^{\rm qc}$.
In order to gain statistics, source operators are inserted at three different
positions. For $\beta=3.4$, for example, the source positions are
$y_0/a=30, 47$ and $65$.

\begin{figure}[htb]
\centerline{%
  \includegraphics[width=0.5\textwidth]{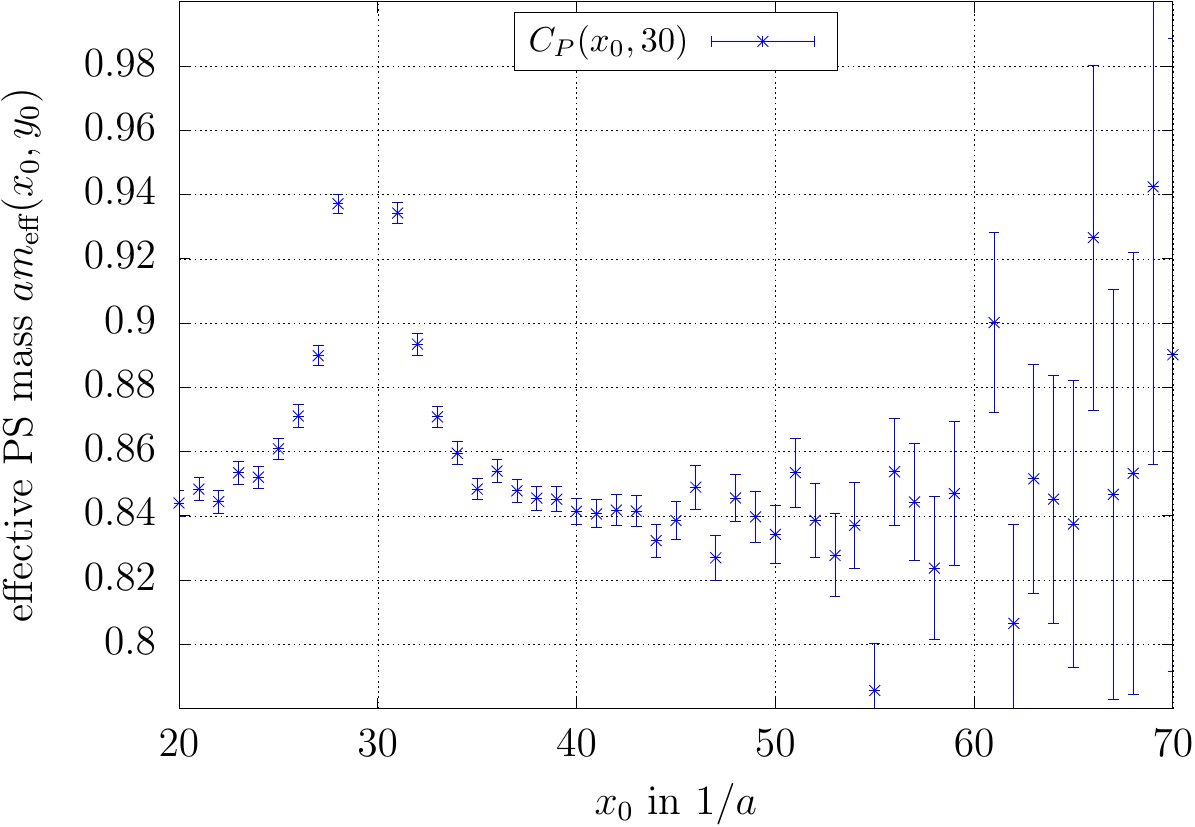}
  \includegraphics[width=0.5\textwidth]{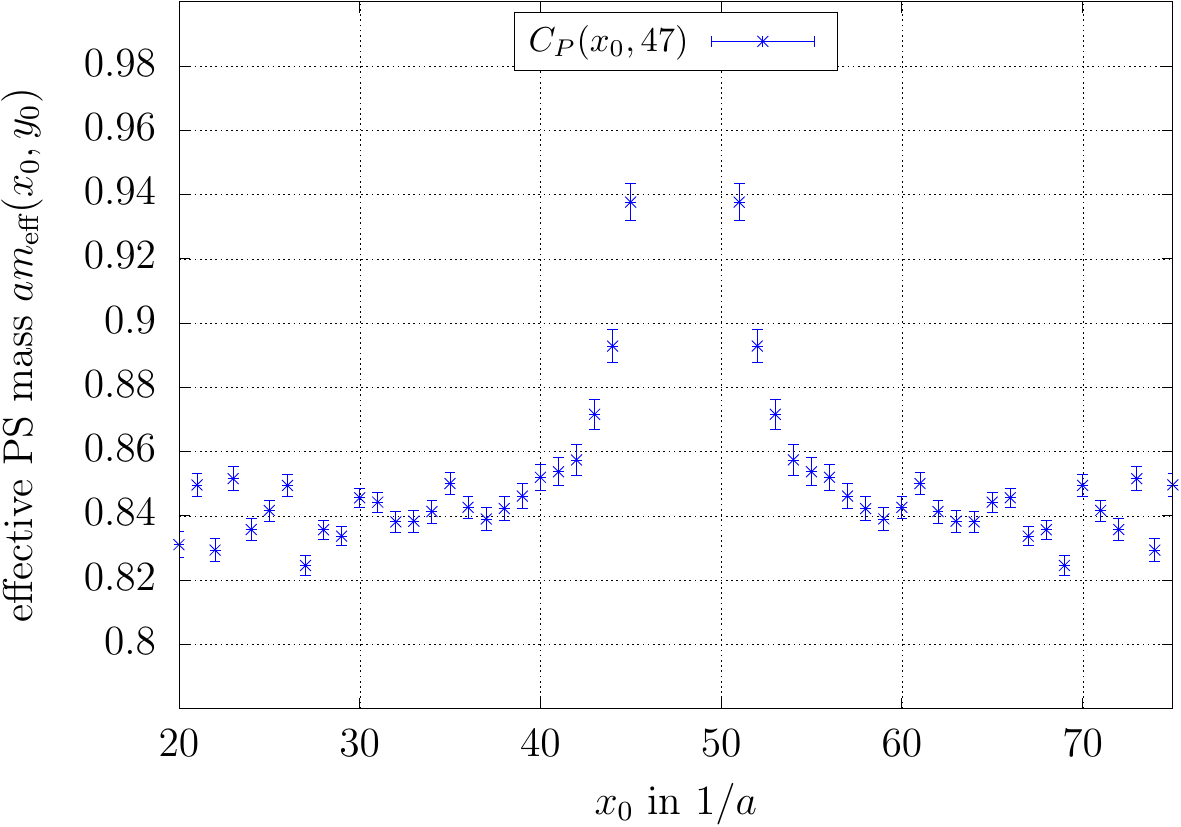}}
  \caption{The effective mass of the pseudoscalar $\rm D$-meson in units of
  $x_0/a$ for the H105 ensemble with $N_t = 96$. The left panel shows the
  effective mass with the source positioned at $x_0/a=30$, while in the right panel
  the source is inserted at $x_0/a=47$.}
  \label{SourcePos3047}
\end{figure}

We exploit time-reversal symmetry of the correlators and average the forward propagating
part of the $C_{\rm X}(x_0,30)$ correlator
with source position at $y_0/a=30$ with the backwards propagating part of the
$C_{\rm X}(x_0,65)$ correlator, while for the $C_{\rm X}(x_0,47)$
correlator we average the forward and backward propagating parts. This yields
four correlators, two for both the axial and the pseudoscalar case. In
order to determine a region where a one-state fit to the ground state can be
done safely, we first perform a double-exponential fit of the form
\begin{equation}
  C_{\rm X}(x_0,y_0) = c_{\mathrm{X},1} \mathrm{e}^{-m_{\mathrm{D}_{\rm q}} (x_0 - y_0)} + c_{\mathrm{X},2} \mathrm{e}^{- M_{\mathrm{D}_{\rm q}}^\prime (x_0 - y_0)} ,
  \label{doubleExponentialFit}
\end{equation}

\begingroup
\setlength{\columnsep}{11pt}%
\begin{wrapfigure}[18]{r}{0.5\textwidth}
\centering
  \hspace{-2.3mm}
  \begin{tikzpicture}
  \sbox0{\includegraphics[width=0.93\linewidth,interpolate=false]{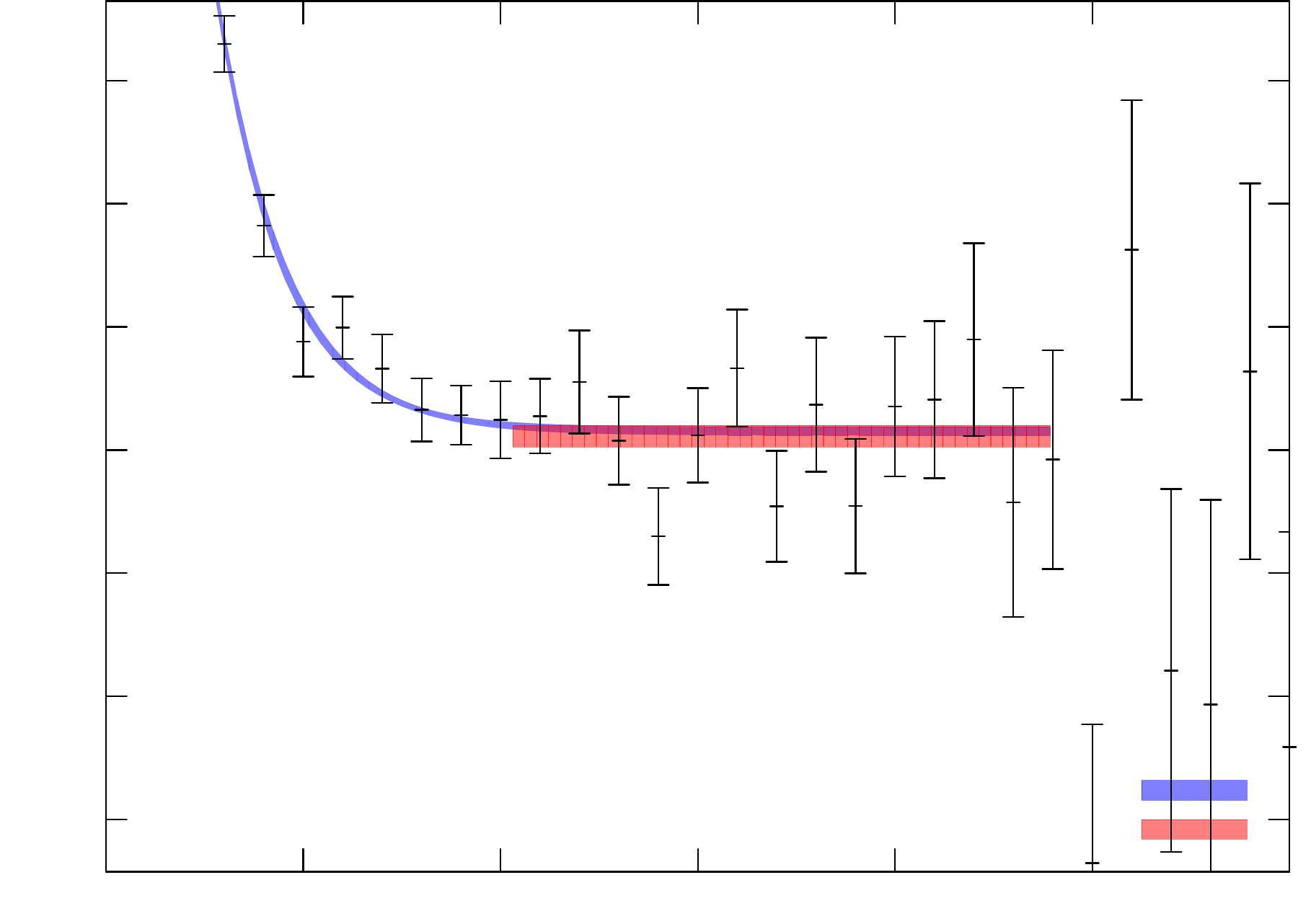}}
  \node[above right,inner sep=0pt] at (0,0)  {\usebox{0}};
  \node at (0.08\wd0,0.01\ht0) {\scalebox{.7}{$0$}};
  \node at (0.231\wd0,0.01\ht0) {\scalebox{.7}{$5$}};
  \node at (0.382\wd0,0.01\ht0) {\scalebox{.7}{$10$}};
  \node at (0.533\wd0,0.01\ht0) {\scalebox{.7}{$15$}};
  \node at (0.684\wd0,0.01\ht0) {\scalebox{.7}{$20$}};
  \node at (0.835\wd0,0.01\ht0) {\scalebox{.7}{$25$}};
  \node at (0.986\wd0,0.01\ht0) {\scalebox{.7}{$30$}};
  \node at (0.53\wd0,-0.07\ht0) {\scalebox{.75}{$(x_0-y_0)/a$}};
  \node[rotate=90] at (-0.05\wd0,0.5\ht0) {\scalebox{.75}{$\text{effective PS mass}\; am_{\rm eff}(x_0,y_0)$}};
  \node at (0.03\wd0,0.115\ht0) {\scalebox{.7}{$0.81$}};
  \node at (0.03\wd0,0.25\ht0) {\scalebox{.7}{$0.82$}};
  \node at (0.03\wd0,0.385\ht0) {\scalebox{.7}{$0.83$}};
  \node at (0.03\wd0,0.515\ht0) {\scalebox{.7}{$0.84$}};
  \node at (0.03\wd0,0.65\ht0) {\scalebox{.7}{$0.85$}};
  \node at (0.03\wd0,0.785\ht0) {\scalebox{.7}{$0.86$}};
  \node at (0.03\wd0,0.915\ht0) {\scalebox{.7}{$0.87$}};
  \node at (0.67\wd0,0.14\ht0) {\scalebox{.6}{double exp. fit, range: 1-24}};
  \node at (0.67\wd0,0.1\ht0) {\scalebox{.6}{single exp. fit, range: 10-24}};
  \end{tikzpicture}
  \vspace{-7mm}
  \caption{Effective mass of the $\rm D$-meson as a function of the
  source-sink separation $(x_0-y_0)/a$ for the H105 ensemble. The double-exponential
  fit determines the starting point $x_0^{\rm min}$ of the
  single-exponential fit to be at $(x_0-y_0)/a=10$ in this case.}
  \label{fittingProcedure}
\end{wrapfigure}
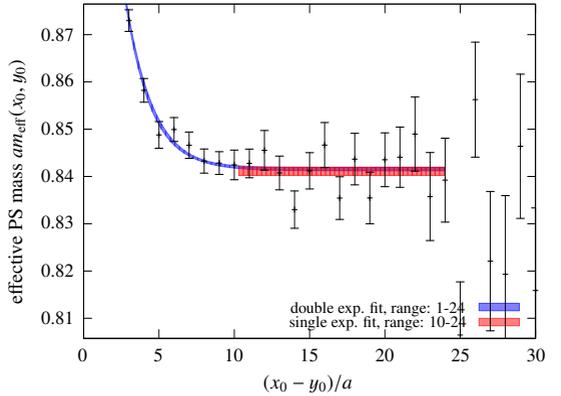
\noindent where the second term represents the first excited state with mass $M_{\mathrm{D}_{\rm q}}^\prime$.
At first, this is done for the pseudoscalar and axial correlators separately.
Since all sources are placed far away from the boundary such that boundary contaminations are expected to be negligible,
$A(y_0)=const.$ (see Eq.~(\ref{eq:corrs_exp})) holds and the amplitudes can be
enforced to be the same for different source positions (e.g., $c_{\mathrm{P},1}$
has the same value for both types of source-position-averaged pseudoscalar correlators).
We vary both the starting point  and the end point of the fit interval in order
to find the optimal fit range, indicated by a minimum of the obtained
$\chi^2$-values. In addition, a careful visual inspection of the fit quality
is performed. Then the point $x_0^{\rm min}$, where the contributions from the
exited states have sufficiently decayed, is given by the criterion

\endgroup
\begin{equation}
  \frac{\left|c_{\mathrm{X},2}\right|^2\,\mathrm{e}^{-M_{\mathrm{D}_{\rm q}}^\prime(x_0^{\rm
        min}-y_0)}}{2M_{\mathrm{D}_{\rm q}}^\prime} < \frac{1}{4}\,\Delta C_{\rm X}(x_0^{\rm
    min},y_0),
    \label{x0min}
\end{equation}

\noindent where $\Delta C_{\rm X}$ denotes the statistical error of the correlator.
Finally, $x_0^{\rm min}$ is taken as the start point for a simultaneous
fit of all four correlators to a single-exponential form:

\begin{equation}
  C_{\rm X}(x_0,y_0) = c_{\mathrm{X},1} \mathrm{e}^{-m_{\mathrm{D}_{\rm q}} (x_0 - y_0)}.
  \label{singleExponentialFit}
\end{equation}

The bare pseudoscalar decay constant $f_{\mathrm{D}_{\rm qc}}^{\rm bare}$ ($\mathrm{q}=\mathrm{l}$
for the $\mathrm{D}$-meson and $\mathrm{q}=\mathrm{s}$ for the $\mathrm{D}_{\rm s}$-meson)
is then given by the ratio
\begin{equation}
  f_{\mathrm{D}_{\rm qc}}^{\rm bare} = \frac{ \sqrt{2} c_{\mathrm{A},1} }{\sqrt{c_{\mathrm{P},1} m_{\mathrm{D}_{\rm q}}}}.
  \label{decayConstant}
\end{equation}

For the end point $x_0^{\rm max}$ of the fit range we choose roughly the point, at which the ratio of the
statistical error and mean of the correlator exceeds $3\,\%$.
Both the double-exponential and the single-exponential fits are illustrated
in Fig.~\ref{fittingProcedure} for the example of the H105 ensemble.

Since our setup does not include a dynamical charm quark, $\kappa_{\rm charm}$
has to be fixed for each ensemble. Based on a subset of statistics, we first estimate $\kappa_{\rm charm}$
from the spin-flavour-averaged $1\mathrm{S}$ mass combination
$M_{\rm X} = \left( 6 m_{\rm D^*}  + 2 m_{\rm D} + 3 m_{\mathrm{D}^*_{\rm s}} + m_{\mathrm{D}_{\rm s}} \right) / 12$
along the $\overline{m}=const.$ line and from the spin-averaged mass combination
$M_{\rm X} = \left( 3 m_{\mathrm{D}^*_{\rm s}} + m_{\mathrm{D}_{\rm s}} \right) / 4$
along the $\hat{m}_{\rm s}=const.$ line. Afterwards,
simulations with full statistics were performed at two values of $\kappa_{\rm charm}$
slightly above and below the estimated value, allowing for an uncertainty
in the lattice spacing $a$ of $\pm 2\,\%$. A linear interpolation of $M_{\rm X}$
in $1/\kappa_{\rm charm}$ to the physical point, given by the central value
of $a$, then fixes the physical $\kappa_{\rm charm}^{\rm (phys)}$. The bare pseudoscalar
decay constants $f_{\mathrm{D}_{(\rm s)}}$ are then interpolated linearly
to this value. Linear interpolations are valid here, since for each ensemble the two chosen
values of $\kappa_{\rm charm}$ are sufficiently close to the target $\kappa_{\rm charm}^{\rm (phys)}$.

\section{Preliminary results}\label{sec-4}
\begin{figure}[h!]
\centerline{%
\includegraphics[width=0.5\textwidth]{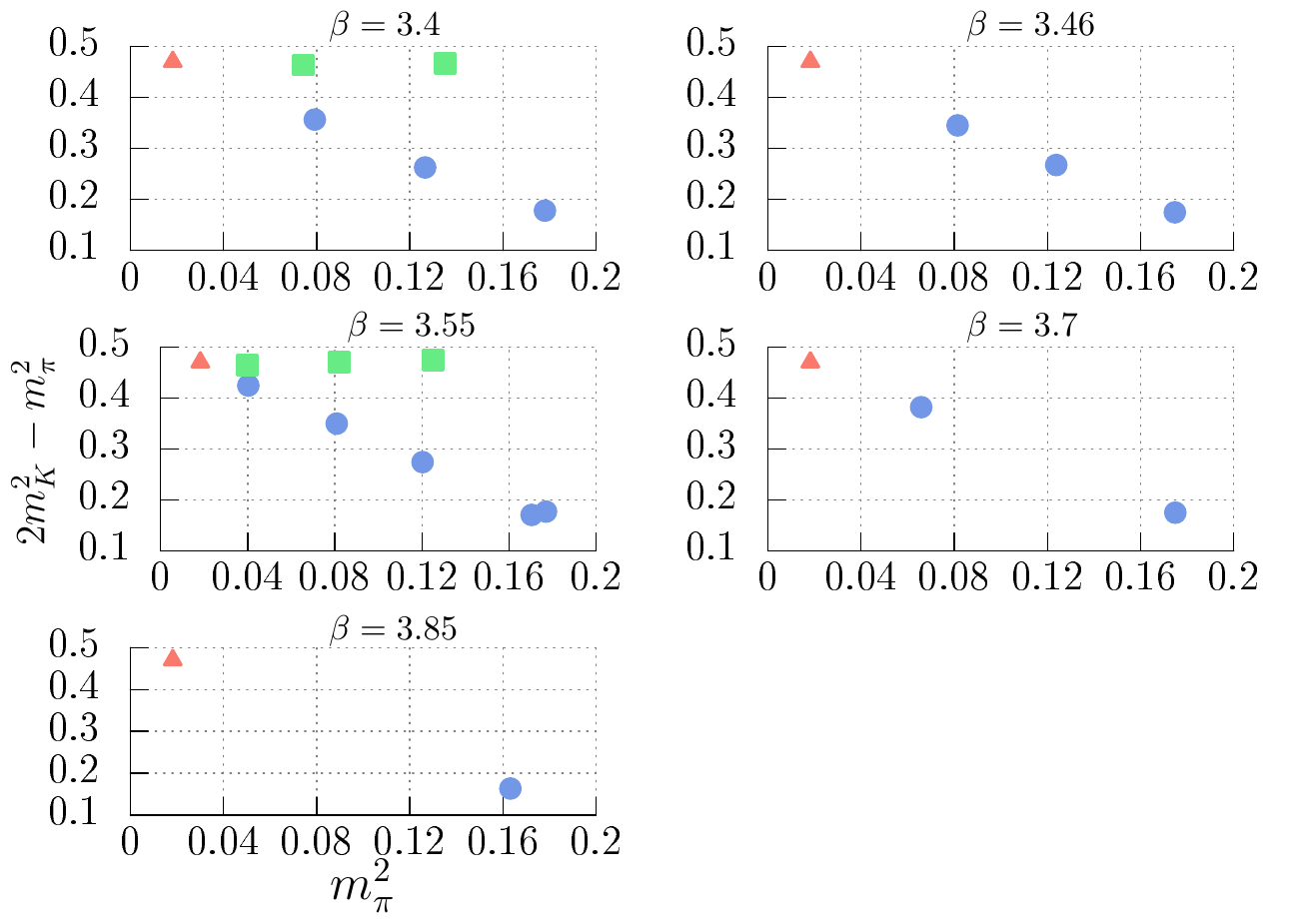}
\includegraphics[width=0.5\textwidth]{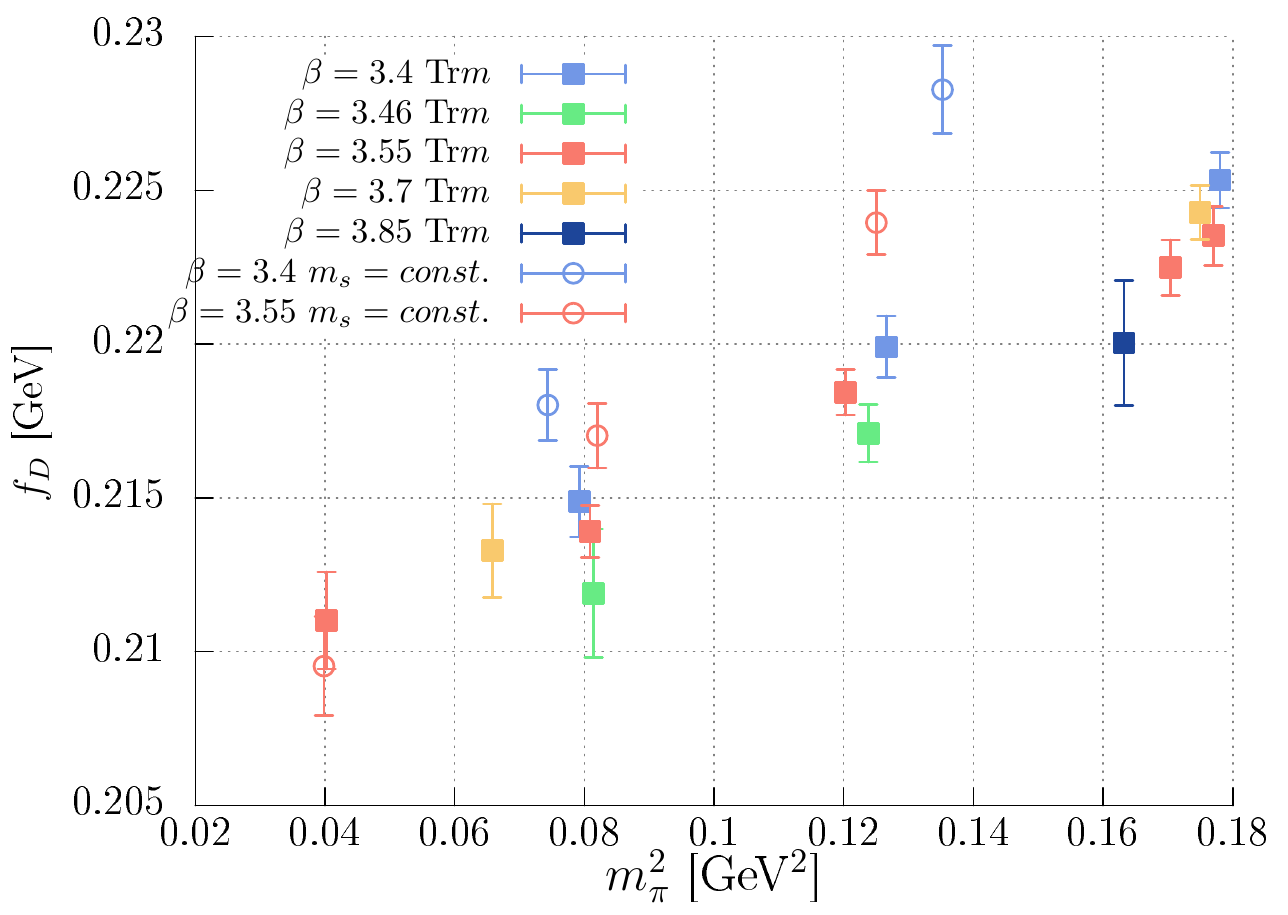}}
    \caption{Ensembles analyzed so far (left panel) and preliminary results
    for $\fD$ as a function of $m_{\pi}^2$ (right panel).}
    \label{ensemblesDmeson}
\end{figure}

\begin{figure}[h!]
\centerline{%
\includegraphics[width=0.5\textwidth]{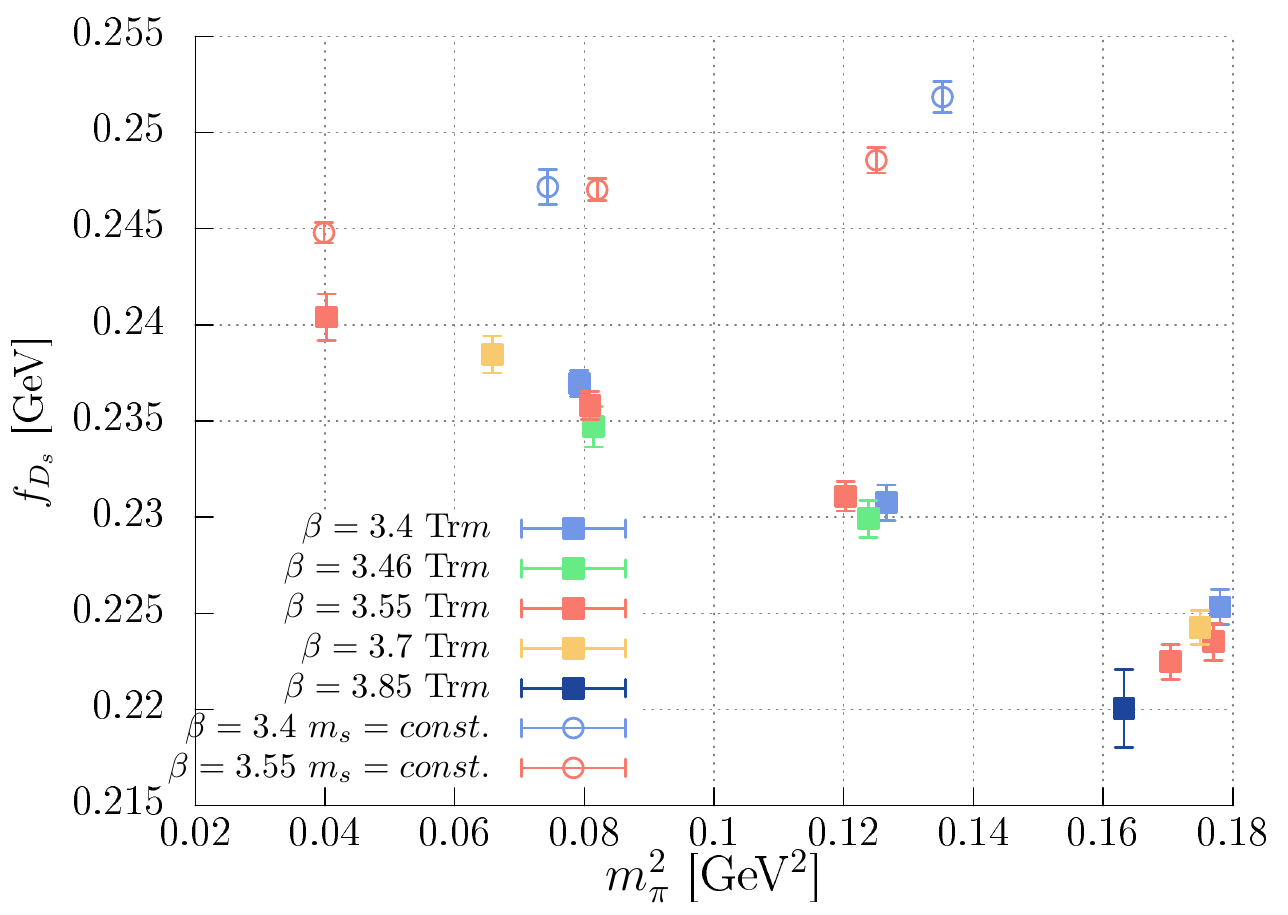}
\includegraphics[width=0.5\textwidth]{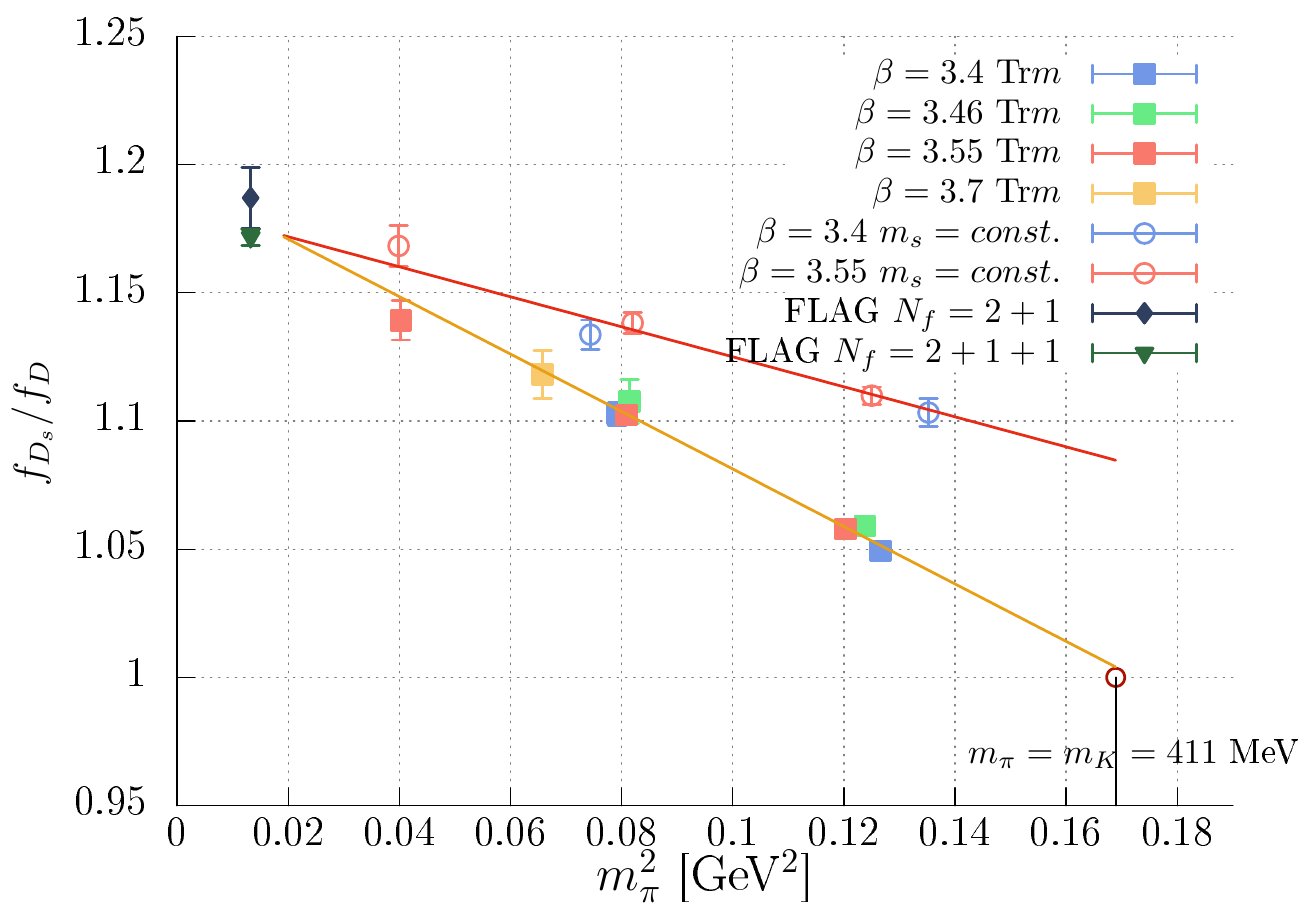}}
    \caption{Preliminary results for $\fDs$ (left panel) and the ratio
    $\fDs / \fD$ as a function of $m_{\pi}^2$ (right panel). The linear fit in the
    right panel was done to guide the eye and does not represent a full
    chiral and continuum extrapolation.}
    \label{DsmesonRatio}
\end{figure}

\noindent
The results described in this section represent the status of the project
at the time of the conference and are summarized in Figs.~\ref{ensemblesDmeson} and~\ref{DsmesonRatio}.
Statistical error analyses are performed using
bootstrap techniques, where the bin size is varied to perform
an extrapolation of the error to infinite bin size, as well as adopting
the so-called $\Gamma$-method, which extracts the statistical errors from the evaluation of
autocorrelation functions~\cite{Wolff:2003sm}. Possible sources of systematic
errors have still to be accounted for in the final analysis.


As the left panel of Fig.~\ref{ensemblesDmeson} shows, the majority of
ensembles at our disposal has been included in the analysis so far. Ensembles for the
$\overline{m}=const.$ line are now available for every lattice spacing
(drawn in blue), while for the $\hat{m}_{\rm s}=const.$ line ensembles have
been included for $\beta=3.4$ and $\beta=3.55$ only (drawn in green). The physical
point in the quark mass plane is indicated by a red triangle. Results for the
pseudoscalar decay constant $\fD$ are presented in the right panel of Fig.~\ref{ensemblesDmeson},
as a function of $m_{\pi}^2$. The left panel of Fig.~\ref{DsmesonRatio} shows the
corresponding results in the case of $\fDs$, while in the right panel the
ratio $\fDs/\fD$ is presented along with the recent FLAG~\cite{Aoki:2016frl} averages for
$N_{\rm f}=2+1$ and $N_{\rm f}=2+1+1$ at the physical point. Combined linear fits in
the right panel do not yet represent a full chiral and continuum extrapolation, but have been
included in order to demonstrate consistency with the FLAG results.
Along the $\overline{m}=const.$ line, the ratio of $\fDs/\fD$ is expected
to be one at the symmetric point $m_{\rm l} = m_{\rm s}$, which is also
in agreement with the linear fits employed. This finding, in conjunction with the
observed absence of large discretization and finite-size effects, leads us to
conclude that these preliminary estimates of the $\mathrm{D}_{\rm (s)}$-meson
decay constants in three-flavour lattice QCD are intermediate promising results.

\section{Outlook}\label{sec-5}
\noindent
In order to improve on the precision of our measurements, we will continue to
increase statistics, as well as to perform a careful analysis of the
statistical errors. Further steps still to be done include gaining a good
control over systematic effects and performing combined chiral and continuum extrapolations.
Furthermore, a cross check of the results via a second method of extracting
the decay constants, utilizing the axial Ward identity (PCAC) quark mass~\cite{Heitger:2013oaa},
is being prepared.
An investigation of the size of the effect from a slight mistuning of the
quark mass trajectories~\cite{Bruno:2014jqa,Bruno:2016plf} is also planned.

\subsection*{Acknowledgments}

We thank Gunnar Bali, Tomasz Korzec, Stefan Schaefer and Rainer Sommer for useful discussions.
This work is supported by the Deutsche Forschungsgemeinschaft (DFG)
through the grants GRK~2149
(\emph{Research Training Group
``Strong and Weak Interactions -- from Hadrons to Dark Matter''},
K.~E. and J.~H.) and the SFB/TRR~55 (S.~C., S.~H. and W.~S.).
We are indebted to our colleagues in CLS for the joint production of the
$\nf=2+1$ gauge configurations.
The authors gratefully acknowledge the Gauss Centre for Supercomputing
e.V. for granting computer time on SuperMUC at the Leibniz
Supercomputing Centre.
Additional simulations were performed on the Regensburg iDataCool cluster
and on the SFB/TRR~55 QPACE~2 and QPACE~B
computers~\cite{Baier:2009yq,Nakamura:2011cd}.
The two-point functions were computed using the
Chroma~\cite{Edwards:2004sx} software package, along with the locally
deflated domain decomposition solver implementation of
openQCD~\cite{openQCD}, the LibHadronAnalysis library and the multigrid
solver implementation of Ref.~\cite{Heybrock:2015kpy}; additional calculations
were carried out using the code based on~\cite{mesons}.


\clearpage
\bibliography{lattice2017}

\end{document}